## ARTICLE

# Coaxial nanowires as plasmon-mediated remote nanosensors

D. Funes-Hernando,[a] M. Peláez-Fernández,[b] D. Winterauer,[c] J.-Y. Mevellec,[a] R. Arenal,[b,d] T. Batten,[c] B. Humbert[a] and J.L. Duvail[a]

This study reports on the plasmon-mediated remote Raman sensing promoted by specially designed coaxial nanowires. This unusual geometry for Raman study is based on the separation, by several micrometres, of the excitation laser spot, on one tip of the nanowire, and the Raman detection at the other tip. The very weak efficiency of Raman emission makes it challenging in a remote configuration. For the proof-of-concept, we designed coaxial nanowires consisting in a gold core to propagate the surface plasmon polaritons and a Raman-emitting shell of poly(3,4-ethylene-dioxythiophene). The success of the fabrication was demonstrated by correlating, for the same single nanowire, a morphological analysis by electron microscopy and Raman spectroscopy analysis. Importantly for probing remote-Raman effect, the original hard template-based process allows to control the location of the polymer shell all along the nanowire, or only close to one or the two nanowire tips. Such all-in-one single nanowires could have applications in the remote detection of photo-degradable substances and for exploring 1D nanosources for integrated photonic and plasmonic systems.

## Introduction

The coupling of plasmonic nanostructures with other materials has been much investigated over the last decade.[1–7] It was motivated mainly by exploiting the localized surface plasmon resonance (LSPR)[1,3,4,7–10], the local heating due to hyperthermia[2] or the near-field absorption and scattering of metallic nanoparticles[3,5,7]. Among these fascinating phenomena, surface-enhanced Raman spectroscopy (SERS) and plasmon-enhanced fluorescence are of particular interest for applications like SERS sensors[1,4,11] and fluorescence sensors[12]. Alternatively, the use of 1D-like metallic nanostructures opens the way to exploit, in a guided way, the propagative nature of surface plasmon polaritons (SPP), i.e. the coupling of photons and charge density oscillations on a metal surface. The combination of the two above-mentioned LSPR and SPP plasmonic effects in 1D hybrid nanostructures is a recent domain of investigation and it is particularly promising for the remote sensing of different species.[7,13–19] Besides applications in nano-sensors, the remote emission is of great interest for nanophotonics as it combines nanowaveguiding and nanosources.[18,20] Alternatively, coupling such systems with a material exhibiting nonlinear optical properties is of interest for emerging devices in plasmonic–organic hybrid technology.[21]

While the fabrication of hybrid organic-inorganic nanoparticles is a tremendous domain of research, the fabrication of 1D-like hybrid nanostructures with high aspect ratio is more challenging. Among different strategies, template and electrospinning methods have been much used to develop long nanowires and nanofibers of many materials.[22] Concerning hybrid metal - polymer nanowires, they have been prepared by different techniques: nano-lithography[11], electrospinning[23–25], self-assembly/chemical reaction[26–28], soft template[29] or hard template[30–37].

In order to achieve a 1D plasmon-mediated remote nanosensor, it is required to get a SPP propagating along the nanowire and a closely located photoactive species. This promotes a coaxial design to get all-in-one single nanowires.

In this scope, we designed coaxial nanowires made of a plasmonic gold core and a conjugated polymer shell of poly(3,4-ethylene-dioxythiophene) (PEDOT) as a Raman emitter, referred as Au@PEDOT nanowires. Coaxial nanowires have been reported previously but their structural and morphological features do not meet the requirements for promoting both SPP and LSPR effects.[38,39] An original hard-template process has been developed that can be exploited for fabricating other metal – (semi)conducting coaxial systems. The success of their fabrication was demonstrated by correlating, for the same single nanowires, the morphological and compositional studies by electron microscopy and the study by Raman spectroscopy. For these Au@PEDOT nanowires, an improvement of the PEDOT supramolecular structure was evidenced, stronger than for pure PEDOT nanowires. This effect has been attributed to the extremely confined electropolymerization within the 5 to 15 nm gaps between the gold nanowire core and the alumina pore surface. Such coaxial nanowires were exploited to demonstrate the proof-of-concept of a remote Raman emission. This was

a. Institut des Matériaux Jean Rouxel (IMN), UMR 6502 and Université de Nantes, 2 rue de la Houssinière, 44322 Nantes, France. E-mail: daniel.funes@cnrs-imn.fr, jean-luc.duvail@univ-nantes.fr
b. Laboratorio de Microscopías Avanzadas, Instituto de Nanociencia de Aragon, Universidad de Zaragoza, c/ Mariano Esquillor Edificio I+D, 50018 Zaragoza, Spain
c. Renishaw plc, New Mills, Wotton-under-Edge, GL12 8JR, United Kingdom
d. ARAID Foundation, 50018 Zaragoza, Spain

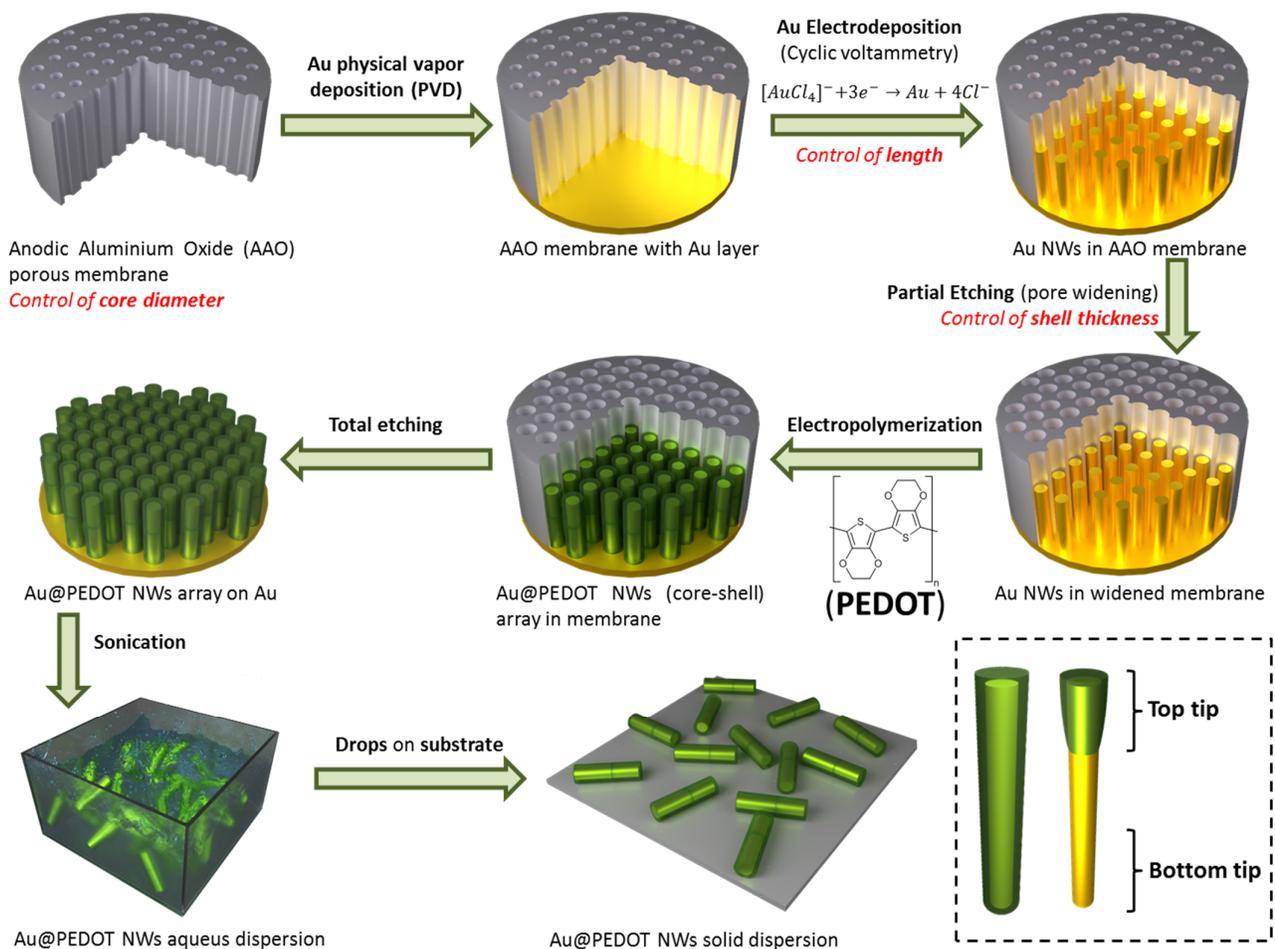

Figure 1 – Scheme of the hard-template process developed to fabricate the coaxial and the asymmetric Au(core)@PEDOT(shell) nanowires.

achieved by exciting one tip of the nanowire with a laser beam and by detecting the Raman signal at the other tip, separated some micrometres. Our study brings original insights to this fewly explored and exploited phenomenon.

## Experimental section

For the fabrication of the Au@PEDOT nanowires, we carried out a templated electrochemical synthesis,[6] using anodic aluminum oxide porous membranes (AAO) as template.[31] Its pore size imposes the diameter of the nanowire Au core. We worked with AAO membranes whose average pore sizes were 22, 42, 75 and 114 nm. Figure 1 summarizes the whole synthesis process.

First of all, a gold layer was sputtered on the barrier layer side of the AAO membrane, typically more organized and with lower pore size dispersity.[§] Then, the metallized AAO membrane was introduced in a homemade electrochemical cell, with the Au layer onto the working electrode and the opposite side in contact with an aqueous solution of $[AuCl_4]^-$, that penetrated the AAO pores. Electroplating of gold was made by cyclic voltammetry (0.00 - 0.75 V vs saturated calomel electrode SCE, scan rate: 50 mV/s) causing the reduction of $[AuCl_4]^-$ ions into $Au^0$ that was first deposited at the bottom of the pores to form nanowires.[40] The length of the nanowires was controlled by changing the number of cycles. In our conditions and for the 114 nm pore size, we measured that 120 cycles produced 12 μm long gold nanowires. Ultrapure water (resistivity > 18.2 MΩ.cm) was used in all the steps of the process.

Then, a pore widening process was performed, as reported in previous works.[6,36,37,41] The $H_3PO_4$ concentration and the etching time make possible to tune the opening size of the pores around the nanowires.[41] The pore wall, in contact with the gold nanowires, was first etched by phosphoric acid due to its richness in anions,[36,37] its higher amorphousness and hydration, and its softening by the localized pH change due to evolution of $H_2$ during the $Au^0$ electrodeposition.[§] We used a 0,9 M phosphoric acid solution in ultrapure water for 90 minutes to get a pore widening of about 5 to 15 nm.

After rinsing the membrane with ultrapure water, an aqueous solution containing the 3,4-ethylenedioxythiophene (EDOT) monomer was dropped to fill the freed space between the Au nanowires and the partially etched AAO pores. Then, the oxidative electropolymerization was carried out at a fixed potential of +0.80 V vs. SCE for 50 s, typically.[34,42,43] Finally, the membrane was treated with concentrated phosphoric acid for 48 hours to totally remove the alumina template. A thin and fragile layer of Au underlying an array of Au@PEDOT nanowires was obtained. To disperse the nanowires in water or ethanol, a brief sonication was applied. The versatility of this process allows to disperse these nanowires on any kind of surface, such as silicon substrates and transmission electron microscopy grids (see ESI for additional information).

In addition, we synthetized asymmetric coaxial Au@PEDOT nanowires with polymer located just at the upper gold tip. This was achieved by shortening the pore widening and the oxidative electropolymerization times. It is emphasized that this kind of asymmetric nanowires was found optimal for investigating the remote Raman excitation, as described in the last part. The lack of PEDOT at the nanowire tip where the laser is focused makes possible to exclude the possibility of any Raman signal excited directly, and not in a remote way.

## Results and discussion

### Characterization of the coaxial nanowires

The morphology and the chemical analysis of the nanowires were investigated by optical microscopy, scanning electron microscopy (SEM) and scanning transmission electron microscopy (STEM). Both arrays and isolated Au(core)@PEDOT(shell) nanowires were studied (Figure 2).

Based on the SEM study of tens of nanowires produced by different synthesis conditions, it was clear on the secondary electron images that the polymer shell - dark grey on images - covered all the gold nanocylinders – white or bright grey - for the case of 22, 44 and 75 nm gold core diameter. When the backscattered electrons were detected, only the gold core appeared while the PEDOT almost disappeared, seemingly a vacuum space in-between the gold cores (see Figure S1). The attribution of this layer to PEDOT was confirmed by scanning transmission electron microscopy – high-angle annular dark-field (STEM-HAADF) imaging (Figure 2(c)) and Raman spectroscopy study (Figure 3).

A systematic study of the PEDOT shell thickness has shown an increase of the thickness with the Au core diameter (Figure 2(b)). We must note that due to the polydispersity of the AAO membranes pore diameter, for the same AAO nominal pore, we could have a range of diameters easily observable on dispersed nanowires. This contrasts with the larger diameter case (nominal pore diameter of 110 nm) where the PEDOT shell covers only the two extremities of the gold core for an equal electropolymerization duration. This observation was confirmed by the SEM study (Figures 2(d) 2(e)) and it was observed that for longer nanowires this unexpected morphology is even more pronounced. The mechanisms responsible for this effect are discussed below.

The linearly polarized Raman spectroscopy study aimed to identify possible molecular and supramolecular ordering of the polymer induced by the ultra-confined synthesis. In addition, plasmon-enhanced effects could take place due to the proximity between the PEDOT and the Au nanowire. It may promote a field-enhanced Raman effect and possibly a SERS effect.[34,38,43,44] Combining unusual supramolecular structure with plasmonic-induced effects is of great interest for exploring the coupling between these effects. Such an investigation is reported in the last part of this manuscript.

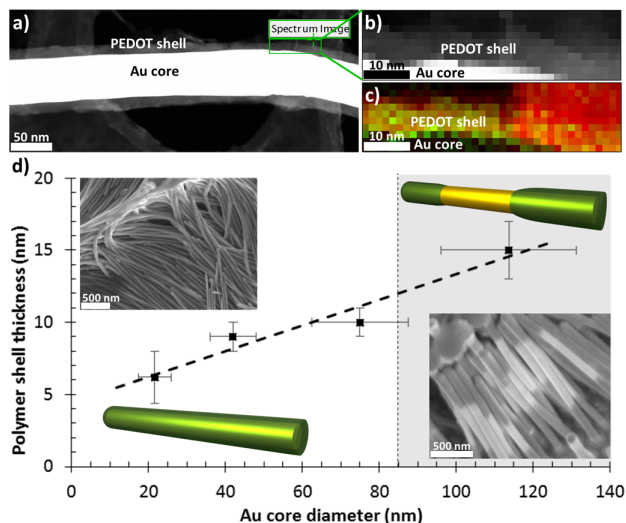

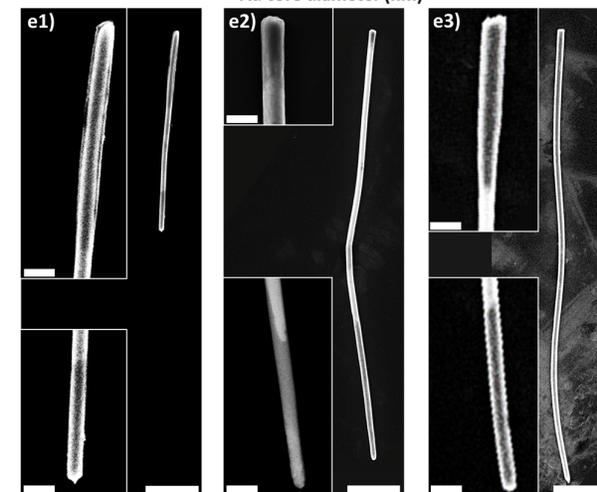

Figure 2 - (a) STEM-HAADF of an Au@PEDOT nanowire. Higher magnification view (b) and STEM-energy electron loss spectroscopy image (c) of the same portion of the nanowire highlighted in (a); carbon is in red and sulfur is in green. (d) Plot of the PEDOT shell thickness versus the Au core diameter for identical synthesis conditions. Inserts show typical SEM images (initial pore diameter: 42 nm left, 114 nm right) and the corresponding artistic schemes of the nanowires with two distinct morphologies: PEDOT (in green) surrounding all the Au (yellow) nanowire for 22, 42 and 75 nm initial pore diameters ; PEDOT surrounding only the extremities of the Au nanowires for 114 nm initial pore diameter. (e1 – e2) SEM images of single nanowires for e1) length L = 4 µm, $t_{synthesis}$ = 50 s, e2) L = 8 µm; $t_{synthesis}$ = 50 s, e3) L = 8.5 µm, $t_{synthesis}$ = 150 s. Scale bars of e1), e2) and e3) are 1 µm (main images) and 200 nm (inserts).

First, a systematic study of the Raman signal was done by mapping single Au@PEDOT and PEDOT nanowires of different diameters. Two configurations of polarization were used to reach information on an eventual ordering effect of the polymer chains‡[24,45,46]: both excitation electric field and backscattering analyzer parallel (perpendicular, respectively) to the nanowire axis. Typical spectra measured for the two polarization configurations on a single coaxial nanowire were compared to the case of a pure PEDOT nanowire (Figure 3). The main bands characteristics of PEDOT were detected. Based on previous theoretical and experimental studies, the most intense band at 1437 - 1441 cm$^{-1}$ is attributed to the C=C symmetric stretching, the band at 1511 cm$^{-1}$ corresponds to the C=C antisymmetric stretching and the C-C stretching band at 1365 cm$^{-1}$.[34] The main band at 1437 cm$^{-1}$ is particularly interesting, because its intensity depends on the length of conjugated segments through resonant effect and on the conjugated segment orientation relative to the excitation polarization. The comparison with the spectra measured for pure PEDOT nanowires electropolymerized in the same conditions (electrolyte, applied voltage) (Figure 3(a)) gives a clear evidence of a strong preferential orientation of the conjugated PEDOT segments perpendicular to the wire axis for the Au@PEDOT nanowire, in contrast with the pure PEDOT nanowire where conjugated segments exhibit a preferential orientation along the nanowire axis. Different groups reported a preferential alignment of the chains for conjugated polymers such as polypyrrole (PPy), polyaniline and PEDOT. It was assigned to a preferential growth of the polymer chains along the nanopore wall.[24,45,47–50] The preferential orientation perpendicular to the nanowire axis in our coaxial nanowires strongly contrasts with this tendency. It has been further investigated acquiring a Raman map along the nanowire, as described below.

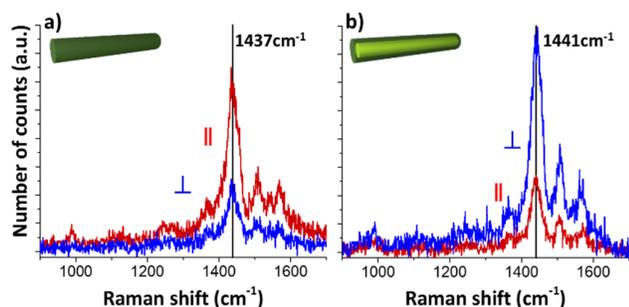

Figure 3 - Polarized Raman spectra measured for (a) a single PEDOT nanowire and (b) a single Au@PEDOT nanowire (synthesized by using the 42 nm pore diameter membrane) with light polarized parallel (red spectrum) and perpendicular (blue spectrum) to the nanowire main axis. ($\lambda_{excitation}$ = 514 nm)

A quantitative analysis of the full-width-at-half-maximum (FWHM) and the I$_{[symetric\ C=C\ stretching]}$/I$_{[antisymmetric\ C=C\ stretching]}$ intensity ratio shows a narrowing of the FWHM and a stronger intensity ratio for the coaxial case (31 cm$^{-1}$; 5.00) than for the pure PEDOT case (38 cm$^{-1}$; 3.55).

Exploiting the conclusions of previous Raman studies on PEDOT nanowires, it suggests a strong molecular and supramolecular structure ordering of the PEDOT shell surrounding the gold core, still more ordered than for the pure PEDOT nanowire.[34,51] This result is discussed later, in view of the variation of the Raman spectra with the excitation wavelength. This preferential orientation of the conjugated PEDOT segments will be exploited for plasmon-mediated remote studies.

Second, a map of the Raman signal at different excitation wavelengths was measured on a 95 nm diameter Au nanowire with PEDOT preferentially grown at the two ends. This Raman map was correlated with the morphology of the nanowire. This morphological information has been obtained by SEM using a specific boron-doped silicon substrate with marks and grids to localize the nanowires.

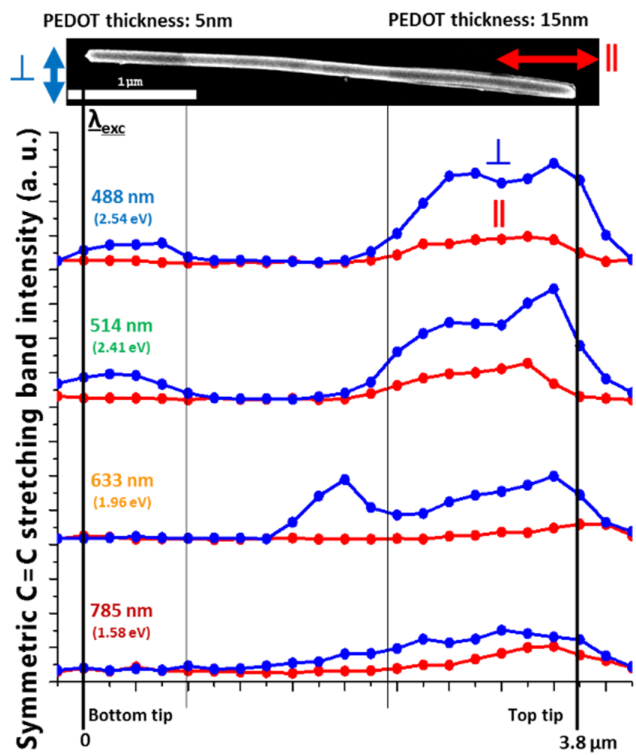

Figure 4 - Scanning electron microscopy (SEM) image and plot of the Raman intensity corresponding to the symmetric C=C stretching of PEDOT along the Au@PEDOT nanowire with PEDOT only at the tips. The intensity mapping is shown for light polarized parallel (red) and perpendicular (blue) to the nanowire main axis (analyzer always parallel to E$_{excitation}$). The diameter of the Au core is 95 nm and its length is 3.8 μm. The table reports the averaged values of the dichroic ratio R (equation 1) for the symmetric C=C stretching band measured along the PEDOT layer located at the bottom or the upper side of the gold nanowire for four excitation wavelengths : 488, 514, 633 and 785 nm.

| R | 488 nm | 514 nm | 633 nm | 785 nm |
|---|---|---|---|---|
| **Bottom tip** | -0.48 | -0.55 | too weak | too weak |
| **Top tip** | -0.52 | -0.41 | -0.65 | -0.30 |

Figure 4 shows the variation of intensity of the Raman C=C symmetric stretching band along the nanowire for the two polarizations (parallel and perpendicular) and for excitation wavelengths at 488, 514, 633 and 785 nm. It has been shown for conjugated polymers that the Raman bands intensity strongly depends on the excitation wavelength in the visible and near infrared range. It was attributed to resonant Raman effects, which depend on the conjugation length and the doping level.[34,52] Indeed, the absorption spectrum of conjugated polymers varies with both the distribution of conjugation lengths and the level of doping.[53] For PEDOT, the absorption spectrum of a strongly doped film electropolymerized under the same conditions (same electrolyte and applied voltage vs SCE) is dominated by the polaronic and bipolaronic bands above 600 nm and in the near infrared domain.[34] Here, the anisotropy of the Raman signal was quantified by calculating the dichroic ratio R, with I$_\parallel$ (I$_\perp$ respectively) the intensity of the C=C symmetric stretching band for excitation and detection in parallel (perpendicular, respectively) polarization:

$$R = \frac{I_\parallel - I_\perp}{I_\parallel + I_\perp} \qquad (1)$$

In the pure PEDOT nanowire (figure S2), the dichroic ratios increased for the smaller diameters, up to R ≈ 0.3 for d = 35 nm, while no preferential orientation (R ≈ 0) was found for d = 200 nm. For the coaxial nanowire, the corresponding values of the dichroic ratio averaged over the 0.5 - 2 μm long PEDOT shell covering the two nanowire extremities are reported in the table of Figure 4 (see Figure S3 for details). Negative values of R correspond to the perpendicular orientation of the conjugated chains. For the thicker PEDOT shell (10 to 15 nm thick) at the upper side of the nanowire, similar values of R (≈ -0.4 ; -0.5) are measured at $\lambda_{exc}$ = 488 nm (2.54 eV) and 514 nm (2.41 eV). It reaches -0.65 at 633 nm (1.96 eV) but only -0.3 at 785 nm (1.58 eV). In the near infra-red domain (785 nm) probing predominantly the longer conjugated segments, the negative dichroic ratio weaker than for other excitation wavelengths can be attributed to a weaker chain alignment perpendicular to the wire axis.

It is interesting to combine this result with the observation that both 633 and 785 nm excitation lines do not detect any contribution of PEDOT at the bottom side of the nanowire, in contrast with the Raman characterization by the 488 and 514 nm excitation wavelengths, as well as with the SEM and STEM study showing a thinner PEDOT layer (5 to 10 nm). These results strongly suggest that the conjugated segments are shorter when grown at the bottom of the gold nanowire, oppositely located to the electrolyte reservoir. It makes sense, that this part is less accessible to the diffusion of the EDOT monomer, dodecyl sulfate and perchlorate anions from the above reservoir than the upper side. Such a mechanism can also explain the reduced electropolymerization rate (thinner layer), which should result in shorter polymer chains and correspondingly shorter conjugated chains for PEDOT at the bottom side than at the upper side. Ultimately, the PEDOT shell at the upper tip can block the 10-15 nm wide channel surrounding the gold cylinder, thus inhibiting further growth of PEDOT at the bottom and on the lateral size when all the monomers remaining in the closed cavity have been consumed. Another possible growth mechanism to explain the preferential electropolymerization at the extremities of the nanowires deals with a tip effect, i.e. an electric-field enhancement which takes place at the top extremity of a strongly anisotropic conducting object connected to an electrode. The larger the aspect ratio (length/diameter), the stronger the electric field enhancement at the tip. Such a local field enhancement could contribute to the overpotential and then promote a faster oxidation of the monomers and thus a faster electropolymerization on the top extremity, rather than on the lateral size. However, such a tip effect cannot account for the growth initiated at the bottom of the nanowire, because no tip effect takes place due to the proximity of the planar electrode on the bottom side of the porous membrane.

The proposed mechanism, which is predominantly responsible for the electropolymerization of PEDOT and its structural ordering on the confined gold nanocylinders, is schematized in

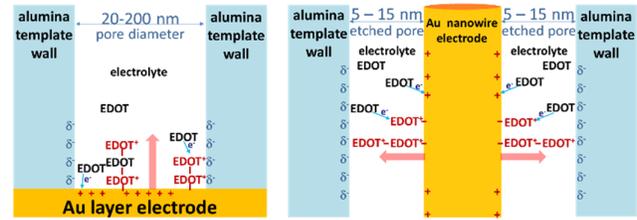

Figure 5 - Simplified schemes of the confined electropolymerization of EDOT in the two geometries. Left: from the Au layer cathode, right: from the Au nanowire.

Figure 5(right), while the predominant mechanism for pure PEDOT nanowires, as previously proposed,[49,54] is reported in Figure 5(left). The resulting improved structure has been demonstrated for different conjugated polymers when synthesized by hard-template methods in nanopores of diameter smaller than 100 nm down to 30 nm.[48,55] For the original geometry of our system, not only the confined space for PEDOT electropolymerization between the gold cylinder and the alumina wall was still reduced down to 5 - 15 nm, but also the geometry of the electrode is modified, replacing the gold electrode at the bottom of the pore by a gold cylinder. These two points can explain the above-mentioned results. While this preferential orientation perpendicular to the nanowire axis was often observed, it is important to mention that a less perpendicular or even a longitudinal orientation was observed in some cases at the bottom tip when exciting with red-infrared lasers (633 nm and 785 nm).

We emphasize that these unique conditions of electrodeposition in an ultra-confined channel and strongly anisotropic nano-electrodes give opportunities to tune standard and unusual mechanisms involved in the growth of 1D-nanomaterials. A direct exploitation is proposed in the evidence of plasmon-mediated remote Raman excitation.

### Remote sensing on coaxial nanowire: a proof-of-concept

The previously described hybrid coaxial nanowires have been exploited for a proof-of-concept of remote-Raman sensing. Different conditions have to be met to get such an effect. First, the excitation of the surface plasmon polariton with a laser can only be achieved, in a non-evanescent configuration, by focusing the laser at one tip of the nanowire to meet the criterion of momentum conservation. SPP along the nanowire cannot be excited if the laser focus does not overlap one tip. Second, the losses of the SPP propagation are strongly wavelength dependent. For gold, it has been shown that SPP can propagate along micrometres for an excitation in the red or near-IR range but it is shortened for smaller wavelengths.[56] Third, when the wire diameter lies in the sub-micrometric range, the plasmonic losses along gold nanowires are expected to be reduced and the formation of Fabry-Perot resonances favoured for the larger diameter.[18,57,58] Among the different newly designed nanowires, we selected the 114 nm diameter ones and the 785 nm excitation to meet the criteria. It is important to note that we exploited nanowires with PEDOT at only the upper tip, as shown on Figure 6(c). Thus, a remote Raman effect is expected when a Raman signal can be collected at the upper tip with a coaxial morphology and when the excitation is done at the opposite (bottom) tip.

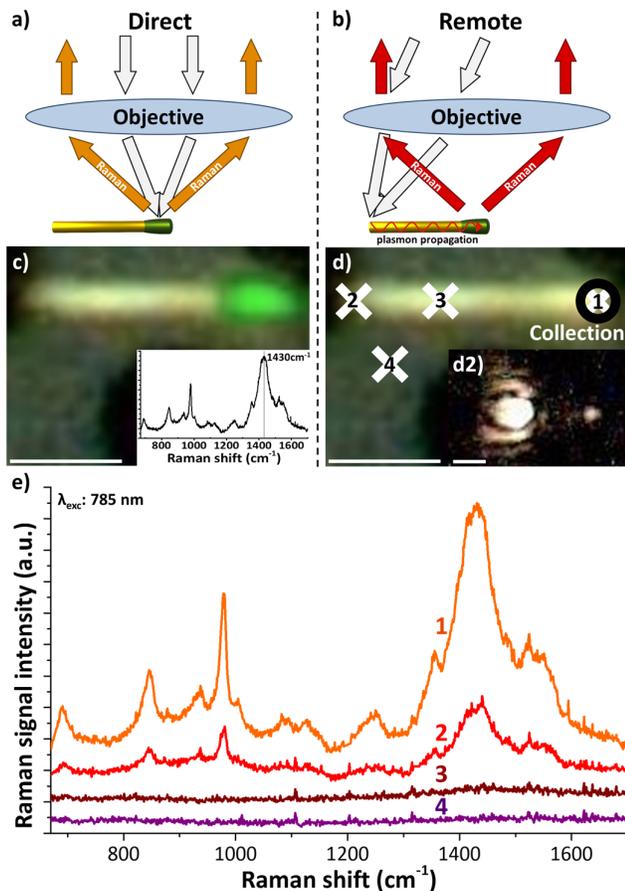

Figure 6 - Schematic view of the optical configuration (laser source, objective, sample and collected light) for measuring the standard (a) and the remote (b) signals. (c) Optical microscopy view (objective x100 LF) with superposed Raman spectroscopy mapping (intensity at 1430 cm⁻¹ appears green) of a single gold nanowire (diameter: 114 nm) partially covered with PEDOT close to the right tip. (d) Four excitation configurations (X symbol labelled 1 to 4) and maintained collection area (O) of the Raman signal plotted in (e) at 785 nm. (d2) Optical microscopy image without back light when exciting in 2 with a 785 nm laser. Scale bars are 2 μm.

The spatially offset set-up was developed in a Renishaw® InVia Raman spectrometer (see Figure 6(b)). The spatially-offset separating the excitation and the Raman signal collection can be precisely controlled onto a single nanowire. For a typical study, four measurements configurations were investigated to prevent from erroneous analysis, as illustrated on the Figure 6(c): (1) excitation and collection at the upper tip (standard direct measurement geometry), (2) excitation precisely at the gold tip and collection at the opposite, i.e. coaxial tip (remote geometry), (3) excitation of the nanowire at 1 or 2 μm from the gold tip onto the nanowire and collection at the opposite, (4) excitation of the $SiO_2$ grid at the same distance (about 4 μm) of the coaxial tip of the nanowire. The collecting area was always the same. The measurement geometries (3) and (4) allowed to discriminate the remote plasmon-mediated effect from a parasitic excitation of PEDOT, coming directly from the laser probe (case (3)) or from the supporting silica grid (case (4)). A fifth measurement in the first geometry was systematically done to check that the PEDOT signal did not suffer any degradation. The irradiation energy was 1.5 J during 500 s. The polarization was always parallel to the nanowire main axis to be in concordance to previous studies on silver nanowires, which have shown that the remote Raman signal has a maximum when the incident polarization is parallel to the nanowire.[17,18,20] Such a methodology was applied to different nanowires and identical results were obtained. Figure 6(e) reports typical spectra measured for the four configurations in a nanowire partially covered by PEDOT on one tip. As expected, the stronger signal was obtained in the first direct configuration, showing the main band around 1420 - 1440 cm⁻¹. For the second configuration (remote geometry), a similar spectrum was measured but with an intensity about 3.5 times smaller, as expected due to the coupling efficiency of the laser with the gold nanowire to excite SPP, and to plasmonic losses when propagating along the nanowire. No signature of PEDOT was detected, neither in the third, nor in the fourth configuration of measurement. These results unambiguously show that the Raman signal detected in the second measurement geometry is due to an excitation of PEDOT by the surface plasmon polaritons after their propagation along the gold nanowire when the excitation is operated at the opposite tip, i.e. remote Raman signal. This study opens the way to investigate longer nanowires and design new architectures for remote nanosensors as long as the development of nanosources.

## Conclusion

Plasmon-mediated remote Raman effect has been evidenced on specially designed coaxial Au@PEDOT nanowires. This coaxial morphology promotes an optimal effect due to the immediate proximity of the Raman scattering material with the gold core, thus benefiting for the excitation of a maximum of the evanescent field coming from the SPP. The opportunity to control the position of PEDOT only at one tip of the nanowire, rather than all along the nanowire or at its two extremities, was exploited to demonstrate unambiguously the remote origin of the Raman signal. The quite strong remote intensity in comparison to the one measured in the direct geometry also confirms the pertinence of our design. In addition, it has been shown that the extremely confined nanoreactors, a 5 to 15 nm thick cylindrical shell where the conjugated polymer is electropolymerized, endorse a strong preferential ordering of the polymer shell. This ordering is another factor contributing to a stronger remote Raman effect.

This original coaxial design including a plasmonic core is particularly attractive for efficient remote Raman sensing of photodegradable organic materials. It could also be exploited to fabricate remote nanosources for nanophotonics, by replacing the Raman emitter with a photoluminescent material.

## Conflicts of interest

There are no conflicts to declare.

## Acknowledgements

This work was supported by the EU Marie Skłodowska-Curie "Enabling Excellence" project 642742.

The STEM-EELS works have been conducted in the Laboratorio de Microscopias Avanzadas (LMA) at the Instituto de Nanociencia de Aragon (INA)-Universidad de Zaragoza (Spain), Spanish ICTS National facility. R.A. gratefully acknowledges the support from the Spanish Ministry of Economy and Competitiveness (MINECO) through project grant MAT2016-79776-P (AEI/FEDER, UE).

We would like to thank Nicolas Stephant for his help in the sample preparation and his guidance during the SEM studies. A special thanks to Eric Gautron and Nicolas Gautier for operating the TEM in preliminary TEM investigation.

Part of this study exploited specially-designed silicon substrates with marks and grids provided by Sebastian Nufer from M-Solv Ltd, UK.

## Notes and references

§ This information was provided by Joseph Huber, InRedox Customer Service and Application Support, in personal communications.

‡ We used polarizers for both the excitation and collection light, their disposition was always in direct correlation: (i) excitation: vertical – collection: vertical (ii) excitation: horizontal – collection: horizontal.

1   B. J. Kennedy, S. Spaeth, M. Dickey and K. T. Carron, *J. Phys. Chem. B*, 1999, **103**, 3640–3646.

2   X. Huang and M. A. El-Sayed, *Alex. J. Med.*, 2011, **47**, 1–9.

3   M. N'Gom, J. Ringnalda, J. F. Mansfield, A. Agarwal, N. Kotov, N. J. Zaluzec and T. B. Norris, *Nano Lett.*, 2008, **8**, 3200–3204.

4   V. Giannini, A. I. Fernández-Domínguez, S. C. Heck and S. A. Maier, *Chem. Rev.*, 2011, **111**, 3888–3912.

5   N. Lagos, M. M. Sigalas and E. Lidorikis, *Appl. Phys. Lett.*, 2011, **99**, 063304.

6   T. Ozel, G. R. Bourret and C. A. Mirkin, *Nat. Nanotechnol.*, 2015, **10**, 319–324.

7   J. K. Day, N. Large, P. Nordlander and N. J. Halas, *Nano Lett.*, 2015, **15**, 1324–1330.

8   T. W. Odom and G. C. Schatz, *Chem. Rev.*, 2011, **111**, 3667–3668.

9   R. Arenal, L. Henrard, L. Roiban, O. Ersen, J. Burgin and M. Treguer-Delapierre, *J. Phys. Chem. C*, 2014, **118**, 25643–25650.

10  H. Y. Feng, F. Luo, R. Arenal, L. Henrard, F. García, G. Armelles and A. Cebollada, *Nanoscale*, 2016, **9**, 37–44.

11  L. Billot, M. Lamy de la Chapelle, A.-S. Grimault, A. Vial, D. Barchiesi, J.-L. Bijeon, P.-M. Adam and P. Royer, *Chem. Phys. Lett.*, 2006, **422**, 303–307.

12  M. Li, S. K. Cushing and N. Wu, *Analyst*, 2015, **140**, 386–406.

13  K. Yu, M. S. Devadas, T. A. Major, S. S. Lo and G. V. Hartland, *J. Phys. Chem. C*, 2014, **118**, 8603–8609.

14  J.-C. Weeber, A. Dereux, C. Girard, J. R. Krenn and J.-P. Goudonnet, *Phys. Rev. B*, 1999, **60**, 9061.

15  J. A. Hutchison, S. P. Centeno, H. Odaka, H. Fukumura, J. Hofkens and H. Uji-i, *Nano Lett.*, 2009, **9**, 995–1001.

16  Y. Fang, H. Wei, F. Hao, P. Nordlander and H. Xu, *Nano Lett.*, 2009, **9**, 2049–2053.

17  Y. Huang, Y. Fang and M. Sun, *J. Phys. Chem. C*, 2011, **115**, 3558–3561.

18  Y. Huang, Y. Fang, Z. Zhang, L. Zhu and M. Sun, *Light Sci. Appl.*, 2014, **3**, e199.

19  J. de Torres, P. Ferrand, G. Colas des Francs and J. Wenger, *ACS Nano*, 2016, **10**, 3968–3976.

20  Z. Li, K. Bao, Y. Fang, Y. Huang, P. Nordlander and H. Xu, *Nano Lett.*, 2010, **10**, 1831–1835.

21  W. Heni, Y. Kutuvantavida, C. Haffner, H. Zwickel, C. Kieninger, S. Wolf, M. Lauermann, Y. Fedoryshyn, A. F. Tillack, L. E. Johnson, D. L. Elder, B. H. Robinson, W. Freude, C. Koos, J. Leuthold and L. R. Dalton, *ACS Photonics*, 2017, **4**, 1576–1590.

22  A. Garreau and J.-L. Duvail, *Adv. Opt. Mater.*, 2014, **2**, 1122–1140.

23  A. Arinstein, M. Burman, O. Gendelman and E. Zussman, *Nat. Nanotechnol.*, 2007, **2**, 59–62.

24  L. M. Bellan and H. G. Craighead, *Polymer*, 2008, **49**, 3125–3129.

25  A. Camposeo, I. Greenfeld, F. Tantussi, S. Pagliara, M. Moffa, F. Fuso, M. Allegrini, E. Zussman and D. Pisignano, *Nano Lett.*, 2013, **13**, 5056–5062.

26  Z. Hu, C. Kong, Y. Han, H. Zhao, Y. Yang and H. Wu, *Mater. Lett.*, 2007, **61**, 3931–3934.

27  X. Ye, C. Zheng, J. Chen, Y. Gao and C. B. Murray, *Nano Lett.*, 2013, **13**, 765–771.

28  N. Jiang, L. Shao and J. Wang, *Adv. Mater.*, 2014, **26**, 3282–3289.

29  M. Aldissi, *J. Polym. Sci. Part C Polym. Lett.*, 1989, **27**, 105–110.

30  G. E. Possin, *Rev. Sci. Instrum.*, 1970, **41**, 772–774.

31  L. S. Van Dyke and C. R. Martin, *Langmuir*, 1990, **6**, 1118–1123.

32  C. R. Martin, *Science*, 1994, **266**, 1961–1966.

33  D. Routkevitch, T. Bigioni, M. Moskovits and J. M. Xu, *J. Phys. Chem. A*, 1996, **100**, 14037–14047.

34  J. L. Duvail, P. Retho, S. Garreau, G. Louarn, C. Godon and S. Demoustier-Champagne, *Synth. Met.*, 2002, **131**, 123–128.

35  J. L. Duvail, P. Rétho, V. Fernandez, G. Louarn, P. Molinié and O. Chauvet, *J. Phys. Chem. B*, 2004, **108**, 18552–18556.

36  W. Lee and S.-J. Park, *Chem. Rev.*, 2014, **114**, 7487–7556.

37  L. Wen, R. Xu, Y. Mi and Y. Lei, *Nat. Nanotechnol.*, 2016, **12**, 244–250.

38  G. Lu, C. Li, J. Shen, Z. Chen and G. Shi, *J. Phys. Chem. C*, 2007, **111**, 5926–5931.

39  S. Harish, J. Mathiyarasu and K. L. N. Phani, *Mater. Res. Bull.*, 2009, **44**, 1828–1833.

40  O. Reynes and S. Demoustier-Champagne, *J. Electrochem. Soc.*, 2005, **152**, D130–D135.

41  J. Kohl, M. Fireman and D. M. O'Carroll, *Phys. Rev. B*, 2011, **84**, 235118.

42  N. Sakmeche, S. Aeiyach, J.-J. Aaron, M. Jouini, J. C. Lacroix and P.-C. Lacaze, *Langmuir*, 1999, **15**, 2566–2574.

43  D. H. Park, M. S. Kim and J. Joo, *Chem. Soc. Rev.*, 2010, **39**, 2439.

44  Ü. Dogan, M. Kaya, A. Cihaner and M. Volkan, *Electrochimica Acta*, 2012, **85**, 220–227.

45  W. Liang and C. R. Martin, *J. Am. Chem. Soc.*, 1990, **112**, 9666–9668.

46  Z. Cai, J. Lei, W. Liang, V. Menon and C. R. Martin, *Chem. Mater.*, 1991, **3**, 960–967.

47  S. Pagliara, M. S. Vitiello, A. Camposeo, A. Polini, R. Cingolani, G. Scamarcio and D. Pisignano, *J. Phys. Chem. C*, 2011, **115**, 20399–20405.

48  C. R. Martin, *Acc. Chem. Res.*, 1995, **28**, 61–68.

49  Y. Shirai, S. Takami, S. Lasmono, H. Iwai, T. Chikyow and Y. Wakayama, *J. Polym. Sci. Part B Polym. Phys.*, 2011, **49**, 1762–1768.

50  V. Singh, T. L. Bougher, A. Weathers, Y. Cai, K. Bi, M. T. Pettes, S. A. McMenamin, W. Lv, D. P. Resler, T. R. Gattuso, D. H. Altman, K. H. Sandhage, L. Shi, A. Henry and B. A. Cola, *Nat. Nanotechnol.*, 2014, **9**, 384–390.

51  M. Akimoto, Y. Furukawa, H. Takeuchi, I. Harada, Y. Soma and M. Soma, *Synth. Met.*, 1986, **15**, 353–360.

52  M. Łapkowski and A. Proń, *Synth. Met.*, 2000, **110**, 79–83.
53  J. L. Brédas and R. Silbey, *Conjugated Polymers: The Novel Science and Technology of Highly Conducting and Nonlinear Optically Active Materials*, Springer Science & Business Media, 1991.
54  M. Granström and O. Inganäs, *Polym. Pap.*, 1995, **36**, 2867–2872.
55  J. L. Duvail, Y. Long, S. Cuenot, Z. Chen and C. Gu, *Appl. Phys. Lett.*, 2007, **90**, 102114.
56  R. M. Dickson and L. A. Lyon, *J. Phys. Chem. B*, 2000, **104**, 6095–6098.
57  H. Ditlbacher, A. Hohenau, D. Wagner, U. Kreibig, M. Rogers, F. Hofer, F. R. Aussenegg and J. R. Krenn, *Phys. Rev. Lett.*, 2005, **95**, 257403.
58  Z. Li, F. Hao, Y. Huang, Y. Fang, P. Nordlander and H. Xu, *Nano Lett.*, 2009, **9**, 4383–4386.